\documentclass[english]{iopart}
\usepackage[T1]{fontenc}
\usepackage[latin9]{inputenc}
\usepackage{graphicx}
\usepackage{amssymb}

\makeatletter

\providecommand{\tabularnewline}{\\}

\usepackage{iopams}
\usepackage{setstack}
  \@ifundefined{theoremstyle}{\usepackage{amsthm}}{}
  \theoremstyle{plain}
  \newtheorem{thm}{Theorem}[section]
  \theoremstyle{plain}
  \newtheorem{conjecture}[thm]{Conjecture}
  \theoremstyle{plain}
  \newtheorem{prop}[thm]{Proposition}

\usepackage{amsfonts}

\makeatother

\usepackage{babel}

\begin{document}

\title{Exotic Smoothness and Quantum Gravity}

\author{T. Asselmeyer-Maluga }

\address{German Aerospace Center, Berlin and }

\address{Loyola University, New Orleans, LA, USA}

\ead{torsten.asselmeyer-maluga@dlr.de}

\begin{abstract}
Since the first work on exotic smoothness in physics, it was folklore
to assume a direct influence of exotic smoothness to quantum gravity.
Thus, the negative result of Duston \cite{Duston2009} was a surprise.
A closer look into the semi-classical approach uncovered the implicit
assumption of a close connection between geometry and smoothness structure.
But both structures, geometry and smoothness, are independent of each
other.

In this paper we calculate the {}``smoothness structure'' part of
the path integral in quantum gravity assuming that the {}``sum over
geometries'' is already given. For that purpose we use the knot surgery
of Fintushel and Stern applied to the class $E(n)$ of elliptic surfaces.
We mainly focus our attention to the K3 surfaces $E(2)$. Then we
assume that every exotic smoothness structure of the K3 surface can
be generated by knot or link surgery a la Fintushel and Stern. The
results are applied to the calculation of expectation values. Here
we discuss the two observables, volume and Wilson loop, for the construction
of an exotic 4-manifold using the knot $5_{2}$ and the Whitehead
link $Wh$. By using Mostow rigidity, we obtain a topological contribution
to the expectation value of the volume. Furthermore we obtain a justification
of area quantization.
\end{abstract}


\submitto{\CQG}

\maketitle

\section{Introduction}

One of the outstanding problems in physics is the unification of the
quantum and general relativity theory. In the case of the quantization
of gravity one works on the space of (pseudo-)Riemannian metrics or
better on the space of connections (Levi-Civita or more general).
The diffeomorphism group of the corresponding manifold is the gauge
group of this theory. Knowing more about the structure of this group
means knowing more about quantum gravity. An explicit expression for
this correspondence is given by General Relativity Theory (GRT). In
this theory we start with a manifold, choose a smooth structure and
write down the field equation. The beginning and the end of this procedure
can be motivated by physics but the choice of the smoothness structure
is not obvious. The reason for this ambiguity is given by the existence
of distinct, i.e. non-diffeomorphic, smoothness structures (exotic
smoothness) in dimension four (see \cite{Asselmeyer2007} for an overview).
The first examples of exotic 7-spheres were discovered by Milnor in
1956 \cite{Mil:56}. Then it was shown that there are only a finite
number of smoothness structures in dimensions greater than $4$ \cite{KerMil:63,KirSie:77}.
In contrast there are countable infinite many (distinct) exotic smoothness
structures for some compact 4-manifolds (see the knot surgery below)
and uncountable infinite many for some non-compact manifolds (see
\cite{BizEtn:98} in case of $M\times{\mathbb{R}}$ with $M$ a compact,
close 3-manifold, or the overview in \cite{GomSti:1999}) including
the exotic ${\mathbb{R}}^{4}$\cite{Gom:85,FreTay:86,Tau:87}. Usually
such a variety of structures has a meaning in physics. The first relation
between quantum field theory and exotic smoothness was found by Witten
\cite{Wit:88.1}. He constructed a topological field theory having
the Donaldson polynomials as ground state. S{\l{}}adkowski \cite{Sla:96,Sla:96b,Sla:96c}
discussed the influence of differential structures on the algebra
$C(M)$ of functions over the manifold $M$ with methods known as
non-commutative geometry. Especially in \cite{Sla:96b,Sla:96c} he
stated a remarkable connection between the spectra of differential
operators and differential structures. For example, the eta-invariant
$\eta({\mathbb{R}P}^{4},g,\phi)$ (measuring the asymmetry of the
spectrum) of a twisted Dirac-operator is different from the eta-invariant
of the exotic ${\mathbb{R}P}^{4}$ for all metrics $g$ and all spin
structures $\phi$ \cite{Sto:88}. Thus the most interesting dimension
for physics has the richest structure. Now the questions are: What
is the physical relevance of the differential structure? How does
the observable or its expectation value depend on the smoothness structure?
In the case of the exotic ${\mathbb{R}}^{4}$ the first question was
discussed by Brans and Randall \cite{BraRan:93} and later by Brans
\cite{Bra:94a,Bra:94b} alone to guess, that exotic smoothness can
be a source of non-standard solutions of Einsteins equation. The author
published an article \cite{Ass:96} to show the influence of the differential
structure to GRT for compact manifolds of simple type. Finally S{\l{}}adkowski
showed in \cite{Sladkowski2001} that the exotic ${\mathbb{R}}^{4}$
can act as the source of the gravitational field. As shown in \cite{AssBra:2002,Sladkowski2009}
the existence of exotic smoothness has a tremendous impact on cosmology,
further explored in \cite{AssRos:07}. Beginning with the work of
Krol \cite{Krol:04a,Krol:04b,Krol:2005,Krol2006,Krol2006a,Krol2010}
on the relation between categorical constructions and exotic smoothness,
the focus in the topic is shifted to show a direct relation between
exotic smoothness and quantum gravity \cite{AsselmeyerKrol2009,AsselmeyerKrol2009a,AsselmeyerKrol2010}. 

In the first section of chapter 10 in \cite{Asselmeyer2007}, the
meaning of exotic smoothness for general relativity was discussed
on general grounds. Since the first papers about exotic smoothness
it was folklore to state an influence of exotic smoothness to the
state sum (or path integral) for quantum gravity. Then the negative
result of the first paper written by Duston \cite{Duston2009} was
a surprise. He was able to calculate semi-classical results using
exotic smoothness. The reason for the negative result is the loose
relation between differential topology and differential geometry for
4-manifolds (in contrast to the strong relation for 3-manifolds known
as the geometrization conjecture of Thurston \cite{Thurston1982}).
But the smoothness structure is rather independent of the geometry
whereas Duston assumed a coupling between geometry and topology because
of its semi-classical approach. Of course, there are some restrictions
induced by the smoothness structure like the existence of a Ricci-flat
metric but nothing more. 

In this paper we use a different approach to overcome this problem
by dividing the measure in the path integral into two parts: the geometrical
part (sum over geometries) and the differential topological part (sum
over exotic smoothness). Both parts are more or less independent of
each other. If one assume the existence of the path integral for the
geometrical part (state sum $Z_{0}$ see below) then the state sum
is changed by exotic smoothness according to (\ref{eq:relation-state-sum}).
The paper is organized as follows. First we will give a physical motivation
for exotic smoothness structures and present the model assumptions.
Then we study the construction of the special class of elliptic surfaces
and the exotic smoothness structures using knot surgery. In section
\ref{sec:action-functional} we discuss the splitting of the action
functional (using the diffeomorphism invariance of the Einstein-Hilbert
action) according to the knot surgery to get the relation (\ref{eq:relation-actions-3}).
Then in section \ref{sec:Observables} we will discuss the calculation
of the functional integral to get (\ref{eq:relation-state-sum}).
Then we discuss the expectation value of two observables: the volume
and the Wilson loop. The calculation of the examples of the knot $5_{2}$
and the Whitehead link $Wh$ together with some concluding remarks
close the paper.

\section{Physical Motivation and model assumptions}

Einsteins insight that gravity is the manifestation of geometry leads
to a new view on the structure of spacetime. From the mathematical
point of view, spacetime is a smooth 4-manifold endowed with a (smooth)
metric as basic variable for general relativity. Later on, the existence
question for Lorentz structure and causality problems (see Hawking
and Ellis \cite{HawEll:94}) gave further restrictions on the 4-manifold:
causality implies noncompactness, Lorentz structure needs a codimension-1
foliation. Usually, one starts with a globally foliated, noncompact
4-manifold $\Sigma\times\mathbb{R}$ fulfilling all restrictions where
$\Sigma$ is a smooth 3-manifold representing the spatial part. But
other noncompact 4-manifolds are also possible, i.e. it is enough
to assume a noncompact, smooth 4-manifold endowed with a codimension-1
foliation. 

All these restrictions on the representation of spacetime by the manifold
concept are clearly motivated by physical questions. Among the properties
there is one distinguished element: the smoothness. Usually one assumes
a smooth, unique atlas of charts covering the manifold where the smoothness
is induced by the unique smooth structure on $\mathbb{R}$. But as
discussed in the introduction, that is not the full story. Even in
dimension 4, there are an infinity of possible other smoothness structures
(i.e. a smooth atlas) non-diffeomorphic to each other.

If two manifolds are homeomorphic but non-diffeomorphic, they are
\textbf{exotic} to each other. The smoothness structure is called
an \textbf{exotic smoothness structure}.

The implications for physics are obvious because we rely on the smooth
calculus to formulate field theories. Thus different smoothness structures
have to represent different physical situations leading to different
measurable results. But it should be stressed that \emph{exotic smoothness
is not exotic physics!} Exotic smoothness is a mathematical possibility
which should be further explored to understand its physical relevance.

In this paper we use a special class of the 4-manifold. Currently,
there are (uncountable) many examples of exotic, noncompact 4-manifolds
which are hard to describe. Thus we will restrict to the class of
compact 4-manifolds where we have powerful invariants like Seiberg-Witten
invariants \cite{Wit:94SW,Akb:96} or Donaldson polynomials \cite{Don:90,DonKro:90}
to distinguish between different smoothness structures. Secondly we
will not discuss the definition of the path integral and all problems
connected with renormalization, definiteness etc. Third, we don't
discuss the Euclidean and/or Lorentzian signature of the metric. Clearly
the compactness of the 4-manifold implies the Euclidean metric on
the 4-manifold but smoothness question are independent of the metric.
Thus, without loss of generality one can remove one point from the
compact 4-manifold to get a noncompact 4-manifold with Lorentz signature.
Forth, we use the knot surgery of Fintushel and Stern \cite{FiSt:96}
to construct the exotic smoothness structure. This approach assumes
a special class of 4-manifolds ({}``complicated-enough'') which
contains the class of elliptic surfaces among them the important K3
surface. Thus we summarize the assumptions

\begin{enumerate}
\item The 4-manifold is the compact and simple-connected elliptic surface
$E(n)$ where $E(2)$ is the K3 surface.
\item All problems with the definition and calculation of the path integral
are ignored.
\item The signature of the metric is not discussed, i.e. the calculations
are the same for Euclidean or Lorentzian metric.
\item The exotic smoothness structures of the 4-manifold $E(n)$ are constructed
by Fintushel-Stern knot surgery.
\end{enumerate}

\section{Elliptic surfaces and exotic smoothness}

In this section we will give some information about elliptic surfaces
and its construction. First we will give a short overview of the construction
of elliptic surfaces and a special class of elliptic surfaces denoted
by $E(n)$ in the literature. Then we present the construction of
exotic $E(n)$ by using the knot surgery of Fintushel and Stern.

\subsection{Elliptic surfaces \label{sec:log} }

A \emph{complex surfaces} $S$ is a 2-dimensional complex manifold
which is compact and connected. A special complex surface is the \emph{elliptic
surface}, i.e. a complex surfaces $S$ together with a map $\pi:S\to C$
($C$ complex curve, i.e. Riemannian surface), so that for nearly
every point $p\in S$ the reversed map $F=\pi^{-1}(p)$ is an elliptic
curve i.e. a torus%
\footnote{We denote this map $\pi$ as elliptic fibration. Thus every complex
surface which is equipped with a elliptic fibration is an elliptic
surface.%
}. Now we will construct the special class of elliptic surfaces $E(n)$.

The first step is the construction of $E(1)$ by the unfolding of
singularities for two cubic polynomials intersecting each other. The
resulting manifold $E(1)$ is the manifold ${\mathbb{C}P}^{2}\#9\overline{\mathbb{C}P}^{2}$
but equipped with an elliptic fibration. Then we use the method of
fiber sum to produce the surfaces $E(n)$ for every number $n\in\mathbb{N}$.
For that purpose we cut out a neighborhood $N(F)$ of one fiber $\pi^{-1}(p)$
of $E(1)$. Now we sew together two copies of $E(1)\setminus N(F)$
along the boundary of $N(F)$ to get $E(2)$ i.e we define the fiber
sum $E(2)=E(1)\#_{f}E(1)$. Especially we note that $E(2)$ is also
known as \emph{K3-surface} widely used in physics. Thus we get the
recursive definition $E(n)=E(n-1)\#_{f}E(1)$. The details of the
construction can be found in the paper \cite{Gom:91.2}.

\subsection{Knot surgery and exotic elliptic surface}

The main technique to construct an exotic elliptic surface was introduced
by Fintushel and Stern \cite{FiSt:96}, called \emph{knot surgery}.
In short, given a simple-connected, compact 4-manifold $M$ with an
embedded torus $T^{2}$ (having special properties, see below), cut
out $M\setminus N(T^{2})$ a neighborhood $N(T^{2})=D^{2}\times T^{2}$
of the torus and glue in $S^{1}\times S^{3}\setminus N(K)$, with
the knot complement $S^{3}\setminus N(K)$ (see appendix \ref{sec:Knot-complement}).
Thus the construction depends on a knot, i.e. an embedding of the
circle $S^{1}$ into $\mathbb{R}^{3}$ or $S^{3}$. Then we obtain
the new 4-manifold\[
M_{K}=(M\setminus N(T))\cup_{T^{3}}(S^{1}\times(S^{3}\setminus N(K)))\]
from a given 4-manifold $M$ by gluing $M\setminus N(T^{2})$ and
$S^{1}\times S^{3}\setminus N(K)$ along the common boundary, the
3-torus $T^{3}$. It is the remarkable result of Fintushel and Stern
\cite{FiSt:96} for a non-trivial knot $K$ that $M_{K}$ is non-diffeomorphic
to $M$. We remark that the construction can be easily generalized
to links, i.e. the embedding of the disjoint union of circles $S^{1}\sqcup\cdots\sqcup S^{1}$
into $\mathbb{R}^{3}$ or $S^{3}$. The reader not interested in the
details of the construction can now jump to the next section.

The precise definition can be given in the following way for elliptic
surface: Let $\pi\colon S\to C$ be an elliptic surface and $\pi^{-1}(t)=F$
a smooth fiber $(t\in C)$. As usual, $N(F)$ denotes a neighborhood
of the regular fiber $F$ in $S$ (which is diffeomorphic to $D^{2}\times T^{2}$).
Deleting $N(F)$ from $S$ to get a manifold $S\setminus N(F)$ with
boundary $\partial(S\setminus N(F))=\partial(D^{2}\times F)=S^{1}\times F=T^{3}$,
the 3-torus. Then we take the 4-manifold $S^{1}\times(S^{3}\setminus N(K))$,
$K$ a knot, with boundary $\partial(S^{1}\times(S^{3}\setminus N(K)))=T^{3}$
and regluing it along the common boundary $T^{3}$. The resulting
4-manifold\[
S_{K}=(S\setminus N(F))\cup_{T^{3}}(S^{1}\times(S^{3}\setminus N(K)))\]
is obtained from $S$ via knot surgery using the knot $K$. The regular
fiber $F$ in the elliptic surface $S$ has two properties which are
essential for the whole construction:

\begin{enumerate}
\item In a larger neighborhood $N_{c}(F)$ of the regular fiber $F$ there
is a cups fiber $c$, i.e. an embedded 2-sphere of self-intersection
$0$ with a single nonlocally flat point whose neighborhood is the
cone over the right-hand trefoil knot.
\item The complement $S\setminus F$ of the regular fiber is simple-connected
$\pi_{1}(S\setminus F)=1$.
\end{enumerate}
Then as shown in \cite{FiSt:96}, $S_{K}$ is not diffeomorphic to
$S$. The whole procedure can be generalized to any 4-manifold with
an embedded torus of self-intersection $0$ in a neighborhood of a
cusp. 

Before we proceed with the physical interpretation, we will discuss
the question when two exotic $S_{K}$ and $S_{K'}$ for two knots
$K,K'$ are diffeomorphic to each other. Currently there are two invariants
to distinguish non-diffeomorphic smoothness structures: Donaldson
polynomials and Seiberg-Witten invariants. Fintushel and Stern \cite{FiSt:96}
calculated the invariants for $S_{K}$ and $S_{K'}$ to show that
$S_{K}$ differs from $S_{K'}$ if the Alexander polynomials of the
two knots differ. Unfortunately the invariants are not complete. Thus
we cannot say anything about $S_{K}$ and $S_{K'}$ for two knots
with the same Alexander polynomial. But Fintushel and Stern \cite{FinSter:1999,FinSter:2002}
constructed counterexamples of two knots $K,K'$ with the same Alexander
polynomial but with different $S_{K}$ and $S_{K'}$. Furthermore
Akbulut \cite{Akb:99} showed that the knot $K$ and its mirror $\bar{K}$
induce diffeomorphic 4-manifold $S_{K}=S_{\bar{K}}$.

\section{The action functional\label{sec:action-functional}}

In this section we will discuss the Einstein-Hilbert action functional
of the exotic 4-manifold. The main part of our argumentation is additional
contribution to the action functional coming from exotic smoothness.
Here it is enough to consider the exotic smoothness generated by knot
surgery.

We start with an elliptic surface, the K3 surface $M=E(2)$. In this
paper we consider the Einstein-Hilbert action\begin{equation}
S_{EH}(g)=\intop_{M}R\sqrt{g}\: d^{4}x\label{eq:EH-action}\end{equation}
and fix the Ricci-flat metric $g$ as solution of the vacuum field
equations. Now we study the effect to vary the differential structure
by a knot surgery. This procedure effected only a submanifold $N(T^{2})\subset M$
and we consider a decomposition of the 4-manifold\[
M=(M\setminus N(T^{2}))\cup_{T^{3}}N(T^{2})\]
with $N(T^{2})=D^{2}\times T^{2}$ leading to a sum in the action
\[
S_{EH}(M)=\intop_{M\setminus N(T^{2})}R\sqrt{g}\: d^{4}x+\intop_{N(T^{2})}R\sqrt{g}\: d^{4}x\quad.\]
Because of diffeomorphism invariance of the Einstein-Hilbert action,
this decomposition don't depend on the concrete realization with respect
to any coordinate system. Now we construct a new smoothness structure
$M_{K}$ \[
M_{K}=(M\setminus N(T^{2}))\cup_{T^{3}}(S^{1}\times(S^{3}\setminus N(K)))\]
by using a knot $K$, i.e. an embedding $K:S^{1}\to S^{3}$. Then
the 4-manifold $M\setminus N(T^{2})$ with boundary a 3-torus $T^{3}$
appears in both 4-manifolds $M$ and $M_{K}$. Thus we can fix it
and its action\[
S_{EH}(M\setminus N(T^{2}))=\intop_{M\setminus N(T^{2})}R\sqrt{g}\, d^{4}x\]
by using a fixed metric $g$ in the interior $int(M\setminus N(T^{2}))$.
Furthermore we can ignore a possible boundary term because the 3-torus
$T^{3}=\partial(M\setminus N(T^{2}))$ is a flat, compact 3-manifold.
Thus we obtain\begin{equation}
S_{EH}(M_{K})=S_{EH}(M\setminus N(T^{2}))+\intop_{S^{1}\times(S^{3}\setminus N(K))}R_{K}\sqrt{g_{K}}\, d^{4}x\label{eq:relation-action-1}\end{equation}
with a metric $g_{K}$ and scalar curvature $R_{K}$ for the 4-manifold
$S^{1}\times(S^{3}\setminus N(K))$. Now we consider the integral\[
S_{EH}(N(T^{2}))=\intop_{N(T^{2})=D^{2}\times T^{2}}R\sqrt{g}\, d^{4}x\]
over $N(T^{2})=D^{2}\times T^{2}$ w.r.t. a suitable product metric.
The torus $T^{2}$ is a flat manifold and the disk $D^{2}$ can be
chosen to embed flat in $N(T^{2})$. Thus, this integral vanishes
$S_{EH}(N(T^{2}))=0$ and \[
S_{EH}(M\setminus N(T^{2}))=S_{EH}(M)\quad.\]
Using this relation and (\ref{eq:relation-action-1}), we obtain the
following relation\begin{equation}
S_{EH}(M_{K})=S_{EH}(M)+\intop_{S^{1}\times(S^{3}\setminus N(K))}R_{K}\sqrt{g_{K}}\, d^{4}x\label{eq:relation-action-knot-surgery}\end{equation}
between the Einstein-Hilbert action on $M$ with and without knot
surgery. Next we will evaluate the integral \[
\intop_{S^{1}\times(S^{3}\setminus N(K))}R_{K}\sqrt{g_{K}}\, d^{4}x\]
by using a product metric $g_{K}$\[
ds^{2}=d\theta^{2}+h_{ik}dx^{i}dx^{k}\]
with periodic coordinate $\theta$ on $S^{1}$ and metric $h_{ik}$
on the knot complement $S^{3}\setminus N(K)$. We are using the ADM
formalism with the lapse $N$ and shift function $N^{i}$ to get a
relation between the 4-dimensional $R$ and the 3-dimensional scalar
curvature $R_{(3)}$ (see \cite{MiThWh:73} (21.86) p. 520)\[
\sqrt{g_{K}}\, R\: d^{4}x=N\sqrt{h}\:\left(R_{(3)}+||n||^{2}((tr\mathbf{K})^{2}-tr\mathbf{K}^{2})\right)d\theta\, d^{3}x\]
with the normal vector $n$ and the extrinsic curvature $\mathbf{K}$.
Without loss of generality, using the product metric in $S^{1}\times(S^{3}\setminus N(K))$
we can embed $S^{3}\setminus N(K)\hookrightarrow S^{1}\times(S^{3}\setminus N(K))$
in such a manner that the extrinsic curvature has a fixed value or
vanishes $\mathbf{K}=0$ (parallel transport of the normal vector).
Then one obtains\[
\intop_{S^{1}\times(S^{3}\setminus N(K))}R_{K}\sqrt{g_{K}}\, d^{4}x=L_{S^{1}}\cdot\intop_{(S^{3}\setminus N(K))}R_{(3)}\sqrt{h}\, N\, d^{3}x\]
with the length $L_{S^{1}}=\intop_{S^{1}}d\theta$ of the circle $S^{1}$.
Thus the integral on the right side is the 3-dimensional Einstein-Hilbert
action. As shown by Witten \cite{Wit:89.2,Wit:89.3,Wit:91.2}, this
action\[
\intop_{(S^{3}\setminus N(K))}R_{(3)}\sqrt{h}\, N\, d^{3}x=L\cdot CS(S^{3}\setminus N(K),\Gamma)\]
 is related to the Chern-Simons action $CS(S^{3}\setminus N(K),\Gamma)$
(defined in the appendix \ref{sec:Chern-Simons-invariant})\[
\intop_{S^{1}\times(S^{3}\setminus N(K))}R_{K}\sqrt{g_{K}}\, d^{4}x=L_{S^{1}}\cdot L\cdot CS(S^{3}\setminus N(K),\Gamma)\]
with respect to the (Levi-Civita) connection $\Gamma$ and a second
length $L$. At least for the class of prime knots, the knot complements
$S^{3}\setminus N(K)$ have constant curvature and the length has
the order of the volume, i.e. $L=\sqrt[3]{Vol(S^{3}\setminus N(K))}$.
Finally we have the relation (\ref{eq:relation-action-1})\begin{equation}
S_{EH}(M_{K})=S_{EH}(M)+L_{S^{1}}\cdot\sqrt[3]{Vol(S^{3}\setminus N(K))}\cdot CS(S^{3}\setminus N(K),\Gamma)\label{eq:relation-actions-2}\end{equation}
as the correction to the action $S_{EH}(M)$ after the knot surgery.
Finally we will write this relation in the usual units\begin{equation}
\frac{1}{\hbar}S_{EH}(M_{K})=\frac{1}{\hbar}S_{EH}(M)+\frac{L_{S^{1}}\cdot\sqrt[3]{Vol(S^{3}\setminus N(K))}}{L_{P}^{2}}\cdot\pi\cdot CS(S^{3}\setminus N(K),\Gamma)\label{eq:relation-actions-3}\end{equation}

\section{The functional integral}

Now we will discuss the (formal) path integral\begin{equation}
Z=\int Dg\ \exp\left(\frac{i}{\hbar}S[g]\right)\label{eq:path-integral-metric}\end{equation}
and its conjectured dependence on the choice of the smoothness structure.
In the following we will using frames $e$ instead of the metric $g$.
Furthermore we will ignore all problems (ill-definiteness, singularities
etc.) of the path integral approach. Then instead of (\ref{eq:path-integral-metric})
we have\[
Z=\int De\:\exp\left(\frac{i}{\hbar}S_{EH}[e,M]\right)\]
with the action\[
S_{EH}[e,M]=\intop_{M}tr(e\wedge e\wedge R)\]
where $e$ is a 1-form (coframe), $R$ is the curvature 2-form $R$
and $M$ is the 4-manifold. Next we have to discuss the measure $De$
of the path integral. Currently there is no rigorous definition of
this measure and as usual we assume a product measure. Then we have
two possible parts which are more or less independent from each other:

\begin{enumerate}
\item integration $De_{G}$ over geometries 
\item integration $De_{K}$ over different differential structures parametrized
by knots $K$.
\end{enumerate}
We assume that the first integration can be done to get formally\[
Z_{0}(M)=\intop_{Geometries}De_{G}\:\exp\left(\frac{i}{\hbar}S_{EH}[e,M]\right)\]
and we are left with the second integration\[
\intop_{Knots}De_{K}\:\exp\left(\frac{i}{\hbar}S_{EH}[e,M_{K}]\right)\]
by varying the differential structure using the knot surgery. But
the set of different differential structures on a compact 4-manifold
has countable infinite cardinality. Then the integral changes to a
sum\[
\intop_{Knots}De_{K}\:\exp\left(\frac{i}{\hbar}S_{EH}[e,M_{K}]\right)=\sum_{Knots\: K}\exp\left(\frac{i}{\hbar}S_{EH}[e,M_{K}]\right)\]
Now we made the following main conjecture:

\begin{conjecture}
All distinct exotic smoothness structures) of the compact 4-manifold
$M=E(2)$ can be generated by knots and links.\label{con:main-conj}
\end{conjecture}
As mentioned above, Akbulut \cite{Akb:99} showed that the pair of
a knot and its mirror knot (same for links) induces diffeomorphic
smoothness structures. Thus assuming the conjecture \ref{con:main-conj}
and the relation (\ref{eq:relation-actions-3}) for knot surgery,
we obtain for the path integral\begin{equation}
Z=Z_{0}\cdot\sum_{K}\exp\left(i\frac{L_{S^{1}}\cdot\sqrt[3]{Vol(S^{3}\setminus N(K))}}{L_{P}^{2}}\cdot CS(S^{3}\setminus N(K),\Gamma)\right)\label{eq:state-sum-exotic}\end{equation}
where the connection $\Gamma$ is chosen to be the Levi-Civita connection.
Finally exotic smoothness contributes to the state sum of quantum
gravity.

\section{Observables\label{sec:Observables}}

Any consideration of quantum gravity is incomplete without considering
observables and its expectation values. Here we consider two kind
observables:

\begin{enumerate}
\item Volume 
\item holonomy along open and closed path (Wilson loop)
\end{enumerate}
The expectation value for the volume can be calculated via the decomposition
of the 4-manifold $M_{K}$, i.e.\[
M_{K}=(M\setminus N(T^{2}))\cup_{T^{3}}(S^{1}\times(S^{3}\setminus N(K)))\]
leading to\begin{eqnarray*}
Vol(M_{K}) & = & Vol(M)+(Vol(S^{1}\times S^{3}\setminus N(K))-Vol(N(T^{2}))\\
 & = & Vol(M)+(L_{S^{1}}\cdot Vol(S^{3}\setminus N(K))-Vol(N(T^{2}))\end{eqnarray*}
Let \[
\left\langle Vol(M)\right\rangle _{0}=\frac{\int De_{G}\: Vol(M,e_{G})\exp\left(\frac{i}{\hbar}S_{EH}[e,M]\right)}{\int De_{G}\:\exp\left(\frac{i}{\hbar}S_{EH}[e,M]\right)}\]
be the expectation value of the volume w.r.t. the geometry. Using
the linearity of the expectation value, we obtain\[
\left\langle Vol(M_{K})\right\rangle _{0}=\left\langle Vol(M)\right\rangle _{0}+\left\langle Vol(S^{1}\times S^{3}\setminus N(K))\right\rangle _{0}-\left\langle Vol(N(T^{2}))\right\rangle \]
and be choosing a unit length scale for the circle in $S^{1}\times(S^{3}\setminus N(K))$
as well for the torus $T^{2}$in $N(T^{2})$ we have\begin{equation}
\left\langle Vol(M_{K})\right\rangle _{0}=\left\langle Vol(M)\right\rangle _{0}+\left\langle Vol(S^{3}\setminus N(K))\right\rangle -1\:.\label{eq:volume-relation-1}\end{equation}
Thus the volume depends on the volume of the knot complement $S^{3}\setminus N(K)$
only. This result seems not satisfactory because one may choose the
scale of the volume $Vol(S^{3}\setminus N(K))$. Surprisingly, that
is not true! As Thurston \cite{Thu:97} showed there are two classes
of knots: hyperbolic knots and non-hyperbolic knots. The (interior
of the) knot complement of a hyperbolic knot admits a homogeneous
metric of constant negative curvature (normalized to $-1$). In contrast,
the knot complement of non-hyperbolic knots admits no such metric.
As Mostow \cite{Mos:68} showed, every hyperbolic 3-manifold is rigid,
i.e. the volume is a topological invariant. Thus we can scale the
knot complement for non-hyperbolic knots in such a manner that\[
Vol(M_{K})\approx Vol(M)\quad\mbox{non-hyperbolic knot }K\]
whereas \[
Vol(M_{K})=Vol(M)+Vol(S^{3}\setminus N(K))-1\quad\mbox{hyperbolic knot }K\:.\]
Finally we can claim

\begin{prop}
\label{pro:expect-value-volume}The expectation value of the volume
for the 4-manifold $M$ depends on the choice of the smoothness structure
if this structure is generated by knot surgery along a hyperbolic
knot or link.
\end{prop}
It is known that most knots or links are hyperbolic knots or links
among them the figure-8 knot, the Whitehead link and the Borromean
rings. The class of non-hyperbolic knots contains the torus knots
among them the trefoil knot.

Now we discuss the holonomy \[
hol(\gamma,\Gamma)=Tr\left(\exp\left(i\intop_{\gamma}\Gamma\right)\right)\]
along a path $\gamma$ w.r.t. the connection $\Gamma$, as another
possible observable. Then there are two cases:

\begin{enumerate}
\item the path $\gamma$ lies in $M\setminus N(T^{2})$, i.e.it is not affected
by knot surgery, or
\item the path $\gamma$ lies in $N(T^{2})$ as well as in $S^{1}\times(S^{3}\setminus N(K))$.
\end{enumerate}
Obviously, the expectation value of the holonomy for the first case
does not depend on the choice of the differential structure and we
will ignore it. In the second case we choose a path in the knot complement
$S^{3}\setminus N(K)$. As stated above, the action is identical to
the Chern-Simons action. Furthermore it is known that there is an
important subclass among all paths, the closed paths%
\footnote{It is known that the path space is fibered over the closed paths by
constant paths.%
} . Thus we will concentrate on the closed paths, i.e. we consider
the Wilson observable\[
W(\gamma)=Tr\left(\exp\left(i\intop_{\gamma}\Gamma\right)\right)\]
with the expectation value\begin{equation}
\left\langle W(\gamma)\right\rangle =\frac{\int D\Gamma\: W(\gamma)\exp\left(i\,\ell\cdot CS(S^{3}\setminus N(K),\Gamma\right)}{\int D\Gamma\:\exp\left(i\,\ell\cdot CS(S^{3}\setminus N(K),\Gamma\right)}\label{eq:knot-inv}\end{equation}
w.r.t. the setting\[
L_{S^{1}}\cdot\sqrt[3]{Vol(S^{3}\setminus N(K))}=L_{P}^{2}\cdot\ell\:.\]
The Wilson loop depends on the connection only. Thus we switch to
the path integral over the connections instead using the frames. As
discussed by Witten in the landmark paper \cite{Wit:89}, the expectation
value $\left\langle W(\gamma)\right\rangle $ for a knot $\gamma$
is a knot invariant depending on $\ell$, now called the colored Jones
polynomial. If $\gamma$ is the unknot then $\left\langle W(\gamma)\right\rangle $
is an invariant of the 3-manifold in which the unknot embeds. Reshtetikhin
and Turaev \cite{ResTur:91} constructed this invariant (the RT-invariant)
using quantum groups. Thus,

\begin{prop}
\label{pro:expect-value-Wilson-loop}The expectation value for the
Wilson loop $\left\langle W(\gamma)\right\rangle $ is a knot invariant
of the knot $\gamma$ in $S^{3}\setminus N(K)$. The knot $\gamma$
represents a non-contractable loop, i.e. an element in $\pi_{1}(S^{3}\setminus N(K))$.
If $\gamma$ is the unknot then \textup{$\left\langle W(\gamma)\right\rangle $}
is the RT-invariant.
\end{prop}
But that is not the full story. First consider a closed, contractable
curve $\gamma$ in $N(T^{2})=D^{2}\times T^{2}=S^{1}\times(D^{2}\times S^{1})$
representing a knot. The knot surgery changed $N(T^{2})$ to $S^{1}\times S^{3}\setminus N(K)$
but the knot $\gamma$ is untouched. Thus, the knot $\gamma$ in $S^{1}\times S^{3}\setminus N(K)$
can be approximated by a knot in $S^{3}$. Then we have the Wilson
loop $W_{0}(\gamma)$ for that special case and obtain \cite{Wit:89}\begin{equation}
\left\langle W_{0}(\gamma)\right\rangle =\mbox{Jones polynomial }J_{\gamma}(q)\label{eq:Wilson-loop-expectation}\end{equation}
where we have \[
q=\exp\left(\frac{2\pi i}{2+\ell}\right)\]
with the variable $\ell$ defined in (\ref{eq:knot-inv}). For this
result we assumed the holonomy group $SU(2)$ or $SL(2,\mathbb{C})$
for the connection, which will be discussed below.

Secondly consider a closed curve, the unknot, in $N(T^{2})=D^{2}\times T^{2}=S^{1}\times(D^{2}\times S^{1})$
going around the non-trivial loop in the solid torus $D^{2}\times S^{1}$.
Again, the knot surgery changed $N(T^{2})$ to $S^{1}\times S^{3}\setminus N(K)$.
Then the unknot in $N(T^{2})$ is the generator of $\pi_{1}(D^{2}\times S^{1})=\mathbb{Z}$
which can be mapped to the group $\pi_{1}(S^{3}\setminus N(K))$,
the so-called knot group. Now we will concentrate on the semi-classical
approach around the classical solution, i.e. the extrema of the Chern-Simons
functional $CS(S^{3}\setminus N(K),\Gamma)$ in (\ref{eq:state-sum-exotic}).
The extrema of the Chern-Simons action are the flat connections (see
appendix \ref{sec:Chern-Simons-invariant}). Then the Wilson loop
$W(\gamma)$ is a homomorphism\[
W(\gamma):\pi_{1}(S^{3}\setminus N(K))\to G\]
 from the fundamental group $\pi_{1}(S^{3}\setminus N(K))$ into the
structure group $G$ of the connection (usually called gauge group).
From the physical point of view one can argue that the connection
$\Gamma$ is induced by a 4-dimensional connection in $S^{1}\times S^{3}\setminus N(K)$.
Thus one has $SO(3,1)$ or $SO(4)$ (remember we don't discuss the
signature). Interestingly this setting is supported by mathematics
too. Above we divide the knots $K$ into two classes: hyperbolic knots
(= knot complement $S^{3}\setminus N(K)$ with hyperbolic geometry)
and non-hyperbolic knots. According to Thurston \cite{Thurston1982}
(see Scott \cite{Scott1983} for an overview) a hyperbolic geometry
is given by the homomorphism\begin{equation}
\pi_{1}(S^{3}\setminus N(K))\to SO(3,1)=PSL(2,\mathbb{C})\label{eq:hom-hyperbolic-geom}\end{equation}
into the Lorentz group isomorphic to the projective, linear group
as isometry group $Isom(\mathbb{H}^{3})$ of the 3-dimensional hyperbolic
space $\mathbb{H}^{3}$. Thus a natural choice of the structure group
for the connection in the knot complement of a hyperbolic knot is
the Lorentz group!

Next we remark that the structure group gauges via conjugation the
Wilson loop. Denote by $Hom(\pi_{1}(S^{3}\setminus N(K)),SO(3,1))$
the set of homomorphisms (\ref{eq:hom-hyperbolic-geom}) and by \[
Rep(S^{3}\setminus N(K),SO(3,1))=Hom(\pi_{1}(S^{3}\setminus N(K)),SO(3,1))/SO(3,1)\]
the representation variety of gauge invariant, flat connections. Usually
this variety has singularities but as shown in \cite{CulSha:83} one
can construct a substitute, the character variety, which has the structure
of a manifold. This variety (as shown by Gukov \cite{Gukov:2005})
represents the classical solutions. Turaev \cite{Turaev1989,Turaev1991}
introduced a deformation quantization procedure to construct a quantum
version of $Rep(S^{3}\setminus N(K),SO(3,1))$. It works in the neighborhood
of boundary, i.e. in $\partial(S^{3}\setminus N(K))\times[0,1]$.
Then one arrives at the so-called skein space (see \cite{AsselmeyerKrol2010}
for a discussion). Then the expectation value of Wilson loop observables
are given by integrals over some skein space. We will come back to
this point in a forthcoming paper.

\section{Examples}

In this section we will present some concrete calculations based on
the relation (\ref{eq:state-sum-exotic}), i.e. we consider the relation\[
\frac{Z(M_{K})}{Z_{0}(M_{K})}=\exp\left(i\,\frac{L_{S^{1}}\cdot\sqrt[3]{Vol(S^{3}\setminus N(K))}}{L_{P}^{2}}\cdot CS(S^{3}\setminus N(K),\Gamma)\right)\]
for a specific exotic 4-manifold $M_{K}$ (knot surgery of $M$ by
knot $K$) with the Levi-Civita connection $\Gamma$, representing
the correction from the exotic smoothness on $M$. In the following
we will introduce a scaling parameter\begin{equation}
\frac{L_{S^{1}}\cdot\sqrt[3]{Vol(S^{3}\setminus N(K))}}{L_{P}^{2}}=\ell\label{eq:scaling-parameter}\end{equation}
and the usual units (\ref{eq:relation-actions-3}) to get \begin{equation}
\frac{Z(M_{K})}{Z_{0}(M_{K})}=\exp\left(i\ell\cdot\pi\cdot CS(S^{3}\setminus N(K),\Gamma)\right)\:.\label{eq:relation-state-sum}\end{equation}
Usually the Chern-Simons invariant has many values depending on the
complexity of the fundamental group $\pi_{1}(S^{3}\setminus N(K))$.
As argued in the appendix \ref{sec:Chern-Simons-invariant}, one has
to consider the minimum value of the Chern-Simons invariant to get
the Chern-Simons invariant for the Levi-Civita connection.

\begin{figure}
\includegraphics[scale=0.25]{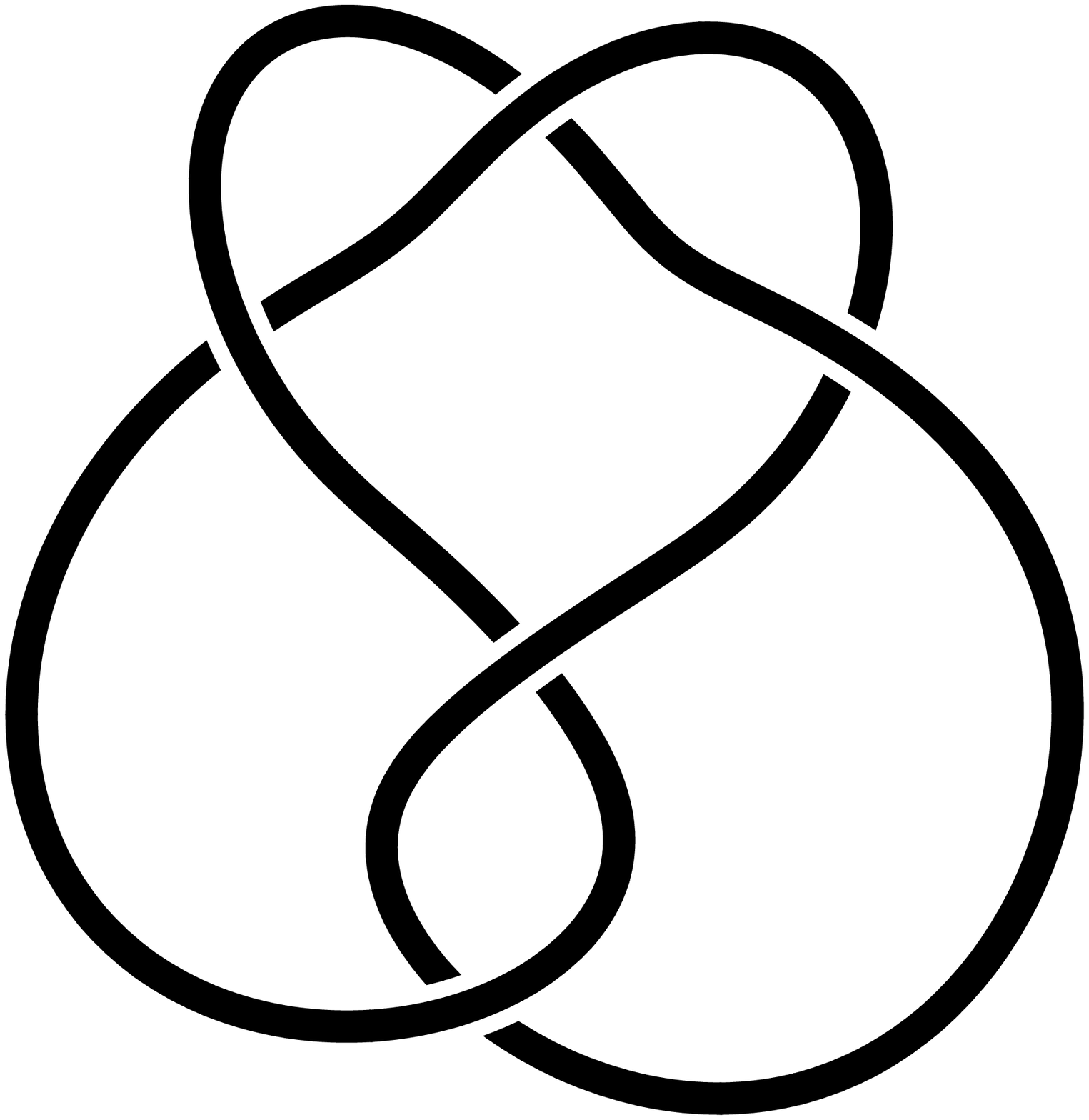}\hfill{} \includegraphics[scale=0.25]{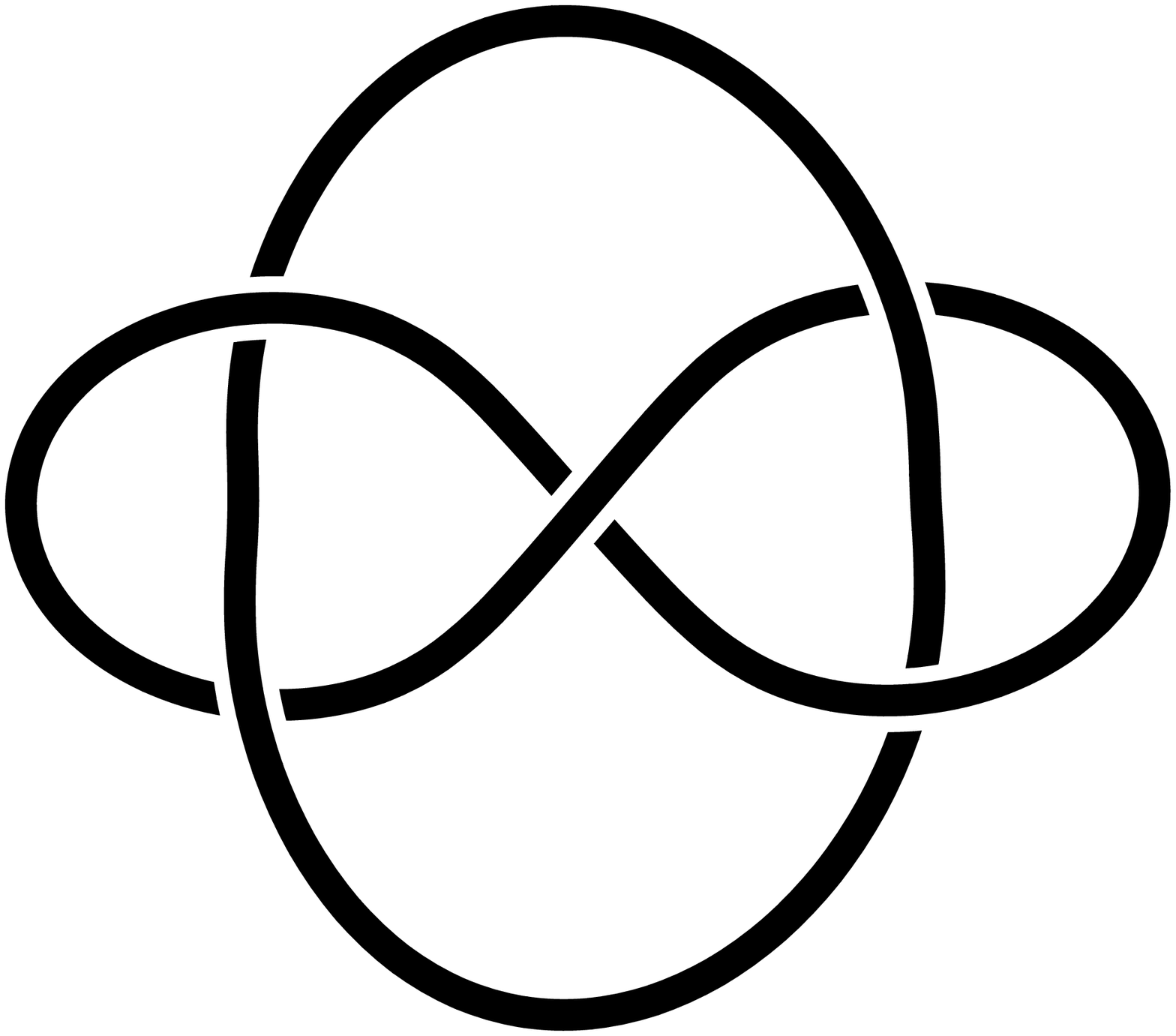}

\caption{Knot $5_{2}$ (left) and Whitehead link $Wh$ (right)\label{fig:Knot-5_2-whitehead}}

\end{figure}

Here we will calculate the observables for two examples where the
value of the volume and of the Chern-Simons invariant is known. The
first example is the hyperbolic knot $5_{2}$ (in Rolfson notation
\cite{Rol:76}) whereas the second example is a two-component, hyperbolic
link, the Whitehead link $Wh$ (see Fig. \ref{fig:Knot-5_2-whitehead}).
 Both complements describe a hyperbolic 3-manifold, i.e. a 3-manifold
with constant negative curvature. The complement $S^{3}\setminus N(5_{2})$
has the boundary $T^{2}$ and $S^{3}\setminus N(Wh)$ has two boundary
components $T^{2}\sqcup T^{2}$. The values of the volume and the
Chern-Simon invariants, calculated via the package SnapPea by J. Weeks,
can be seen in Tab. \ref{tab:volume-and-Chern-Simons}. %
\begin{table}
\caption{volume and Chern-Simons invariant for the knot complement of the knot
$5_{2}$ and the Whitehead link $Wh$\label{tab:volume-and-Chern-Simons}}
\begin{center}\begin{tabular}{|c|c|c|}
\hline 
knot/link & volume & Chern-Simons invariant\tabularnewline
\hline
\hline 
$5_{2}$ & $2.8281220$ & $-3.02412837\bmod\pi^{2}$\tabularnewline
\hline 
$Wh$ & $3.66386$ & $+2.46742\bmod\pi^{2}$\tabularnewline
\hline
\end{tabular}\end{center}
\end{table}
 For the knot $5_{2}$ we obtain the corrections from (\ref{eq:volume-relation-1})
and (\ref{eq:relation-state-sum}) induced by the knot surgery along
$5_{2}$, i.e. for the exotic 4-manifold $M_{5_{2}}$\begin{eqnarray}
\left\langle vol(M_{5_{2}})\right\rangle  & = & \left\langle vol(M)\right\rangle +1.828122\cdot L_{P}^{4}\cdot\ell^{2}\label{eq:knot-5_2-vol}\\
Z(M_{5_{2}}) & = & Z_{0}(M_{5_{2}})\cdot\exp\left(-i\ell\cdot\frac{3.02412837}{\pi}\right)\label{eq:knot-5_2-state-sum}\end{eqnarray}
by assuming the length scale $L_{P}$ for the knot. It is easy to
extend the knot surgery to every link. For the two-component Whitehead
link $Wh$ we cut out two solid tori $N(T^{2}\sqcup T^{2})$ and glue
in $S^{1}\times S^{3}\setminus N(Wh)$. Then we will produce the exotic
4-manifold $M_{Wh}$ with\begin{eqnarray}
\left\langle vol(M_{Wh})\right\rangle  & = & \left\langle vol(M)\right\rangle +2.66386\cdot L_{P}^{4}\cdot\ell^{2}\label{eq:link-Wh-vol}\\
Z(M_{Wh}) & = & Z_{0}(M_{Wh})\cdot\exp\left(i\ell\cdot\frac{2.46742}{\pi}\right)\label{eq:link-Wh-state-sum}\end{eqnarray}
Thus we obtain non-trivial corrections for the observables coming
from exotic smoothness on the 4-manifold $M$.

\section{Conclusion}

In this paper we discuss the influence of exotic smoothness on the
observables, the volume and the Wilson loop, in a model for quantum
gravity. As one would expect from physics, the path integral (or state
sum) as well the expectation values of the observables depend on the
choice of the exotic smoothness structure. For the construction of
the exotic smoothness structure, we use the knot surgery of Fintushel
and Stern \cite{FiSt:96} for a knot. This procedure can be simply
described by substituting a solid (unknotted) torus $D^{2}\times S^{1}$
(seen as the complement $S^{3}\setminus N(S^{1})$ of the unknot)
by a knot complement $S^{3}\setminus N(K)$. Then the result of every
calculation depends strongly on this knot complement. Especially we
found a strong dependence of the volume from the exotic smoothness
for all hyperbolic knots, i.e. knots $K$ where the (interior of the)
complement has a metric of constant, negative curvature (=hyperbolic
structure). In that case, Mostow rigidity \cite{Mos:68} implies the
topological invariance of the volume (see Proposition \ref{pro:expect-value-volume})!
Thus, there is no scaling of the volume so that the contribution to
the expectation value of the volume (\ref{eq:volume-relation-1})
cannot be neglected (see the examples (\ref{eq:knot-5_2-vol},\ref{eq:link-Wh-vol})).
Furthermore, if we consider the the scaling parameter $\ell$ defined
by (\ref{eq:scaling-parameter}) and argue via a consistent quantum
field theory based on the Chern-Simons action then Witten \cite{Wit:91.2}
showed that this parameter must be quantized. Then we have shown that
there is a quantization of the area according to\[
\frac{L_{S^{1}}\cdot\sqrt[3]{Vol(S^{3}\setminus N(K))}}{L_{P}^{2}}\in\mathbb{N}\]
agreeing with results in Loop quantum gravity \cite{RovSmo:95}. \\
\emph{Finally we can support the physically motivated conjecture that
quantum gravity depends on exotic smoothness.}

\ack{}{This work was partly supported by the LASPACE grant. Many thanks
to Carl H. Brans for nearly infinite many discussions about the physics
of exotic 4-manifolds. The author acknowledged for all mathematical
discussions with Duane Randall and Terry Lawson.}

\appendix

\section{Knot complement\label{sec:Knot-complement} }

Let $K:S^{1}\to S^{3}$ be an embedding of the circle into the 3-sphere,
i.e. a knot $K$. We define by $N(K)=D^{2}\times K$ a thickened knot
or a knotted solid torus. The knot complement $S^{3}\setminus N(K)$
results in cutting $N(K)$ off from the 3-sphere $S^{3}$. Then one
obtains a 3-manifold with boundary $\partial(S^{3}\setminus N(K))=T^{2}$.
The properties of the knot complement depend strongly on the properties
of the knot $K$. So, the fundamental group $\pi_{1}(S^{3}\setminus N(K))$
is also denoted as knot group. In contrast, the homology group $H_{1}(S^{3}\setminus N(K))=\mathbb{Z}$
don't depend on the knot.

\section{Chern-Simons invariant\label{sec:Chern-Simons-invariant}}

Let $P$ be a principal $G$ bundle over the 4-manifold $M$ with
$\partial M\not=0$. Furthermore let $A$ be a connection in $P$
with the curvature \[
F_{A}=dA+A\wedge A\]
and Chern class\[
C_{2}=\frac{1}{8\pi^{2}}\int\limits _{M}tr(F_{A}\wedge F_{A})\]
for the classification of the bundle $P$. By using the Stokes theorem
we obtain \[
\int\limits _{M}tr(F_{A}\wedge F_{A})=\int\limits _{\partial M}tr(A\wedge dA+\frac{2}{3}A\wedge A\wedge A)\]
with the Chern-Simons invariant \begin{equation}
CS(A,\partial M)=\frac{1}{8\pi^{2}}\int\limits _{\partial M}tr(A\wedge dA+\frac{2}{3}A\wedge A\wedge A)\:.\label{CS-invariante}\end{equation}
Now we consider the gauge transformation $A\rightarrow g^{-1}Ag+g^{-1}dg$
and obtain\[
CS(g^{-1}Ag+g^{-1}dg,\partial M)=CS(A,\partial M)+k\]
with the winding number \[
k=\frac{1}{24\pi^{2}}\int\limits _{\partial M}(g^{-1}dg)^{3}\in\mathbb{Z}\]
of the map $g:M\rightarrow G$. Thus the expression \[
CS(A,\partial M)\bmod1\]
is an invariant, the Chern-Simons invariant. Now we will calculate
this invariant. For that purpose we consider the functional (\ref{CS-invariante})
and its first variation vanishes \[
\delta CS(A,\partial M)=0\]
because of the topological invariance. Then one obtains the equation
\[
dA+A\wedge A=0\:,\]
i.e. the extrema of the functional are the connections of vanishing
curvature. The set of these connections up to gauge transformations
is equal to the set of homomorphisms $\pi_{1}(\partial M)\rightarrow SU(2)$
up to conjugation. Thus the calculation of the Chern-Simons invariant
reduces to the representation theory of the fundamental group into
$SU(2)$. In \cite{FinSte:90} the authors define a further invariant\[
\tau(\Sigma)=\min\left\{ CS(\alpha)|\:\alpha:\pi_{1}(\Sigma)\rightarrow SU(2)\right\} \]
for the 3-manifold $\Sigma$. This invariants fulfills the relation\[
\tau(\Sigma)=\frac{1}{8\pi^{2}}\int\limits _{\Sigma\times\mathbb{R}}tr(F_{A}\wedge F_{A})\]
which is the minimum of the Yang-Mills action \[
\left|\frac{1}{8\pi^{2}}\int\limits _{\Sigma\times\mathbb{R}}tr(F_{A}\wedge F_{A})\right|\leq\frac{1}{8\pi^{2}}\int\limits _{\Sigma\times\mathbb{R}}tr(F_{A}\wedge*F_{A})\]
i.e. the solutions of the equation $F_{A}=\pm*F_{A}$. Thus the invariant
$\tau(\Sigma)$ of $\Sigma$ corresponds to the self-dual and anti-self-dual
solutions on $\Sigma\times\mathbb{R}$, respectively. Or the invariant
$\tau(\Sigma)$ is the Chern-Simons invariant for the Levi-Civita
connection.

\section*{References}{}

\end{document}